\documentclass{eas}
\usepackage{graphicx}
\TitreGlobal{Tidal effects in stars, planets and disks}
\begin{document}

\title{Tidal dissipation in binary systems} 
\author{Jean-Paul Zahn}\address{LUTH, Observatoire de Paris, F-92195 Meudon, France}

\begin{abstract}
To first approximation, a binary system conserves its angular momentum while it evolves to its state of minimum kinetic energy: circular orbit, all spins aligned, and components rotating in synchronism with the orbital motion. The pace at which this final state is achieved depends on the physical processes that are responsible for the dissipation of the tidal kinetic energy. For stars (or planets) with an outer convection zone, the dominant mechanism identified so far is the viscous dissipation acting on the equilibrium tide. For stars with an outer radiation zone, it is the radiative damping operating on the dynamical tide.

After a brief presentation of the tides, I shall review these physical processes; I shall discuss the uncertainties of their present treatment, describe the latest developments, and compare the theoretical predictions with the observed properties concerning the orbital circularization of close binaries.
\end{abstract}

\maketitle

\section{Introduction}
A fundamental property of closed mechanical systems is that they conserve their total momentum. This is true in particular for binary stars, star-planet(s) systems, whether they possess or not a circumstellar disc, if one can ignore the angular momentum that is carried away by winds and by gravitational waves. Through tidal interaction, kinetic energy and angular momentum are exchanged between the rotation of the components  their orbital motion and the disc. In the absence of such a disc, which is the case that we shall consider here, they evolve due to viscous and radiative dissipation to the state of minimum kinetic energy, in which the orbit is circular, the rotation of both stars is synchronized 
with the orbital motion, and their spin axis are perpendicular to the orbital plane. How rapidly the system tends to that state is determined by the strength of the tidal interaction, and thus by the separation of the two components: the closer the system, the faster its dynamical evolution. But it also depends on the efficiency of the physical processes that are responsible for the dissipation of the kinetic energy. 

Provided these dissipation processes are well enough understood, the observed properties of a binary system can deliver important information on its evolutionary state, on its past history, and even on the conditions of its formation. The first step is thus to identify these physical processes, and it is surprising that this has not been seriously undertaken until the mid-sixties, while tidal theory as such had already reached a high degree of sophistication, starting with the pioneering work of Darwin (1879).  In his classical treatise, Kopal (1959) states from the onset that he is interested only in ``dynamical phenomena which are likely to manifest observable consequences in time intervals of the order of 10 or 100 years, and if so, tidal friction can be safely ignored". 

But stars live much longer than that, and this is why 
 we shall consider here changes in the properties of binary systems that occur over their evolutionary time scale, and in particular the circularization of their orbits, which is both easy to observe and easy to interpret.  We shall deal mainly with binary stars, although much of what follows may be applied also to star-planet systems.

\section{The equilibrium tide}

\begin{figure}[!ht]
\begin{center}
\includegraphics[width=0.7\textwidth] {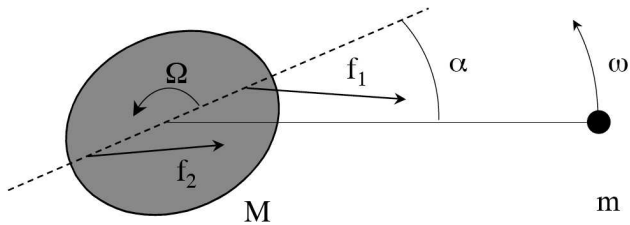}
\end{center}
\caption{Tidal torque. When the considered star rotates faster than the orbital motion ($\Omega > \omega$), its mass distribution is shifted by an angle $\alpha$ from the line joining the centers of the two components, due to the dissipation of kinetic energy. Since the forces applied to the tidal bulges are not equal ($f_1 > f_2$), a torque is exerted on the star, which tends to synchronize its rotation with the orbital motion ($\Omega \rightarrow \omega$).}
\label{tide-torque}
\end{figure}

We begin with the most simple concept: that of the equilibrium tide, where one 
assumes that the star is in hydrostatic equilibrium, and that, in the absence of dissipation mechanisms, it would adjust instantaneously to the perturbing force exerted by its companion.

\subsection{The weak friction approximation}

Let us first establish a rough estimate of the tidal torque.
Take the system depicted in Fig. \ref{tide-torque}, where two stars are separated by the distance $d$, and focus on one of the stars that we shall call the primary. It is not necessarily the most massive one: its mass is $M$ and its radius $R$. The companion star of mass $m$ produces on it two opposite tidal bulges, whose relative elevation $\delta R / R$ equals approximately the ratio of the differential acceleration exerted on the bulges to the surface gravity $g$:
\begin{equation}
{\delta R \over R} \approx  {G m R /d^3 \over GM/R^2}  = {m \over M} 
\left({R \over d} \right)^3 
\end{equation}
($G$ is the gravitational constant).
If the primary star had a constant density, its tidal bulges would have a mass of order $\delta M \approx (\delta R/R) M$; their actual mass is lower, since the surface layers are less dense than the deep interior. These tidal bulges produce a quadrupolar gravity field, which causes the motion of the apsides in an elliptic orbit (see Kopal 1959, and T. Mazeh's contribution to the present volume).

For simplicity, let us assume that the orbit is circular. When the rotation of the star is synchronized with the orbital motion, the tidal bulges are perfectly aligned with the companion star. However, when the rotation is not synchronized, any type of dissipation causes a slight lag of the tidal bulges, and the star then experiences a torque which tends to drag it into synchronism. That torque is easily estimated:
\begin{equation}
\Gamma \approx  (f_2 - f_1) R \sin \alpha \approx - \delta M \left({G m R \over d^3}\right) R \sin \alpha 
= - {G m^2 \over R} \left({R \over d}\right)^6 \sin \alpha,
\end{equation}
where $\alpha$ is the tidal lag angle, and neglecting numerical factors of order unity.

This tidal angle is a function of the lack of synchronism, and it vanishes for $\Omega \rightarrow \omega$, $\Omega$ being the rotation rate and $\omega$ the orbital angular velocity. In the simplest case, which is called the {\it weak friction approximation}, $\alpha$  is proportional to
($\Omega - \omega$). It also depends on the strength of the physical process that is responsible for the dissipation of kinetic energy, which may be measured by its characteristic time $t_{\rm diss}$. This leads us to
\begin{equation}
\alpha = {(\Omega - \omega) \over t_{\rm diss}} {R^3 \over GM},
\label{weak-f}
\end{equation}
where we have rendered $\alpha$ non-dimensional by introducing the most `natural' time, namely the dynamical (or free-fall) time $(GM / R^3)^{-1/2}$.

From this we can deduce the tidal torque 
\begin{equation}
\Gamma = - {(\Omega-\omega) \over  t_{\rm diss}} q^2 M R^2  \left({R \over d}\right)^6,
\label{weak-rot}
\end{equation}
where $q=m/M$ is the mass ratio between secondary and primary components, and $I$ the moment of inertia of the primary star.

The weak friction law (\ref{weak-f}) may be applied to solid, elastic planets, and to first approximation also to fluid bodies, such as giant planets and stars, assuming that the dissipation is of viscous nature, and that the viscosity does not depend on the tidal frequency, hence on $(\Omega-\omega)$. (As we shall see later on, this condition is not necessarily fulfilled.) In that case
the correct expression for the tidal torque, which one derives from the full equations governing the problem,  is precisely of the form given above (Eq.~\ref{weak-rot}).
From it, we may draw the synchronization time $t_{\rm sync}$:
\begin{equation}
{1 \over t_{\rm sync}} = - {1 \over I \Omega} {d (I \Omega) \over dt} =
- {\Gamma \over I \Omega} =   {1 \over  t_{\rm diss}} {(\Omega - \omega) \over \Omega}  q^2 {M R^2 \over I } \left({R \over a}\right)^6 ;
\label{sync}
\end{equation}
here the torque has been averaged here over the orbit, whose semi-major axis is $a$.
The viscous dissipation time is $t_{\rm diss} = R^2/<\nu>$, where $<\nu>$ is a suitable average of the kinematic viscosity, as we shall see later on. 

This expression (\ref{sync}) is strictly valid only for a circular orbit; corrections of order $e^2$ apply when the orbit is elliptic, $e$ being the eccentricity. 
Even better, it is possible to derive an expression for the tidal torque that is valid for any $e$ (Hut 1981).
As a result, the torque averaged over the elliptic orbit no longer vanishes for $\Omega=\omega$, but for 
\begin{equation}
{\Omega \over \omega} = {1 + {15 \over 2} e^2 + {45 \over 2} e^4 + {5 \over 16} e^6 \over
(1 - e^2)^{3/2} (1 + 3 e^2 + {3 \over 8} e^4)} ,
\end{equation}
where  {\it pseudo-synchronization} is achieved (cf. T. Mazeh in this volume). 
\begin{figure}[b]
\begin{center}
\includegraphics[width=0.4\textwidth] {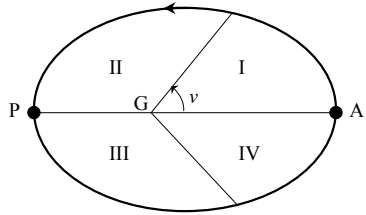}
\end{center}
\caption{Orbital circularization in the weak friction approximation. The figure shows the orbit of the primary in an inertial frame, G being the center of gravity of the system; it has been divided in four sectors,  which have been chosen such that the star spends a quarter of the period in each of them. Its rotation is pseudo-synchronized, i.e. on average it experiences no tidal torque as it moves along the orbit. In sectors II and III, the orbital angular velocity $\omega=dv/dt$ exceeds the rotational angular velocity $\Omega$, hence the torque is accelerating the rotation and therefore is decelerating the orbital motion, since angular momentum is conserved: this will reduce the distance GA of the apastron, and thus acts to decrease the orbital eccentricity. In sectors IV and I, $\Omega > \omega$, the orbital motion is accelerated, the distance GP of the periastron increases, and this works also to diminish the eccentricity.}
\label{circ-schema}
\end{figure}

Since the instantaneous orbital velocity varies along an elliptic orbit, so does also the torque applied to the primary. When the rotation is pseudo-synchronized, that reduces the orbital eccentricity, as illustrated 
in fig.~\ref{circ-schema}. The circularization is governed by
\begin{equation}
{1 \over t_{\rm circ}} = - {d \ln e \over dt} =   {1 \over  t_{\rm diss}}  \left(9 - {11 \over 2} {\Omega \over \omega } \right) q(1+q)   \left({R \over a}\right)^8;
\label{circ}
\end{equation}
the companion star contributes a similar amount. Note that synchronization proceeds much faster that circularization, because the angular momentum of the orbit ($\approx M a^2  \omega$) is in general much larger than that stored in the stars ($I \Omega < M R^2 \Omega$). It was Darwin (1879) who first pointed out that the eccentricity increases when $\Omega / \omega > 18/11$.

All these expressions assume implicitly that the angular velocity is constant throughout the star. This is not necessarily the case, since the tidal torque varies with depth, as we shall see, and therefore it tends to impose a state of differential rotation. The problem then becomes much more intricate, since one has to deal in addition with the transport of angular momentum within the star.

Another difficulty arises because such differential rotation modifies the velocity field induced by the tidal force. This was pointed out by Scharlemann (1982), who took that effect into account; however his treatment is strictly applicable only when the orbit is circular, and it has yet to be extended to elliptic orbits. 

\subsection{Turbulent convection: the most powerful mechanism for tidal dissipation}\label{turbconv}
In stellar interiors, the viscosity due to microscopic processes is very low: it amounts typically to $\nu \approx 10 - 10^3\,$ cm$^2$s$^{-1}$. The (global) viscous timescale $R^2 / \nu$ therefore exceeds the age of the Universe.

But viscosity still plays a major role in those regions of stars and planets that are turbulent. There the kinetic energy of the large scale flow that is induced by the tide cascades down to smaller and smaller scales, until it is dissipated into heat by viscous friction. The force which acts on the tidal flow may then be ascribed to a `turbulent viscosity' of order $\nu_t \approx v \ell$, where $v$ is the r.m.s. vertical velocity of the turbulent eddies, and $\ell$ their vertical mean free path (or mixing-length). The tidal torque is then expressed by an integral of the turbulent viscosity over the whole star, and the dissipation time scales as the global convective turn-over time:
\begin{equation}
{1 \over t_{\rm diss}} = { 6 \lambda_2 \over t_{\rm conv}} \quad \hbox{with} \quad t_{\rm conv}= \left({M R^2 \over L} \right)^{1/3},
\label{tconv}
\end{equation}
where the non-dimensional constant $\lambda_2$ is given by
\begin{equation}
{\lambda_2 \over t_{\rm conv} } = {336 \over 5} \pi \, {R \over M} \int x^8 \rho \nu_t dx ,
\label{lambda2}
\end{equation}
$L$ being the luminosity, $\rho$ the density and $x=r/R$ the normalized radial coordinate.
This expression was established assuming that the whole luminosity is carried by the convective flux.

This dissipation time is very short: $t_{\rm conv}=0.435$ yr in the present Sun, and for this reason turbulent convection is the most powerful dissipation mechanism acting on the equilibrium tide (Zahn 1966b). It works particularly well in stars possessing an outer convection zone, such as solar-type stars. Assuming that the whole heat flux is carried by convection and that the star is fully convective, $\lambda_2 = 0.019 \, \alpha^{4/3}$, with $\alpha$ (not to be confused with the tidal lag introduced above) being the classical mixing-length parameter (Zahn 1989).

Tidal dissipation due to turbulent convection is considerably reduced in stars with a convective core, as it scales as $(r_c/R)^7$ with the radius $r_c$ of that core (Zahn 1966b). Furthermore, in such cores the convective turnover time may exceed the tidal period, and therefore the straightforward definition  of the turbulent viscosity taken above, i.e. $\nu_t \approx v \ell$, can no longer be applied, as we shall see next. 

\subsection{How to deal with fast tides}
\label{fast-tide}
When I encountered that problem in my thesis work, I made the naive assumption that when the convective turnover time $t_{\rm conv} = \ell/v$ exceeds the tidal period $P_{\rm tide}$, the mean free path should be replaced by the distance that turbulent eddies are crossing during, say, half a tidal period. The turbulent viscosity is then given by
\begin{equation}
\nu_t =   v \ell \min [1,  P_{\rm tide}/ t_{\rm conv}] ,
\label{nuz}
\end{equation}
ignoring numerical coefficients of the order of unity (Zahn 1966b). This reduction affects mainly the deepest layers of a convection zone, since the convective turn-over time increases roughly as the 3/2 power of depth.

The same problem was met somewhat later by Goldreich and Nicholson (1977), when they estimated the tidal damping in Jupiter. They remarked that ``though the largest convective eddies move across distances of order $\ell P_{\rm tide}/ t_{\rm conv}$ in a tidal period, they do not exchange momentum with the mean flow on this time scale". Assuming that the Kolmogorov spectrum applies to convective turbulence, they retained in that spectrum only the eddies whose turnover time (or life time) is less than a tidal period; in that case, the turbulent viscosity is reduced to
\begin{equation}
\nu_t =   v \ell \min [1,  (P_{\rm tide}/ t_{\rm conv})^2] .
\label{nugn}
\end{equation}
They concluded that ``tidal interactions between Jupiter and its satellites have played a negligible role in the evolution of the latters'  orbits".

Recently Goodman and Oh (1997) re-discussed the problem, and they proposed yet another scaling, namely
\begin{equation}
\nu_t =   v \ell \min [1,  (P_{\rm tide}/ t_{\rm conv})^{5/3}] ,
\label{nugo}
\end{equation}
but after examining the behavior of a dynamical toy model for convection, they concluded that (\ref{nugn}) was preferable.

The question of which of these prescriptions should be applied is still considered as Achille's heel of tidal theory. One may even question the validity of the very concept of turbulent viscosity, since we know that stratified convection is hardly a diffusive process: the transport of heat and  momentum is partly achieved by long-lived plumes, and it is not easy to predict how these will interact with the large scale tidal flow.

\begin{figure}[!ht]
\begin{center}
\includegraphics[width=.8\textwidth] {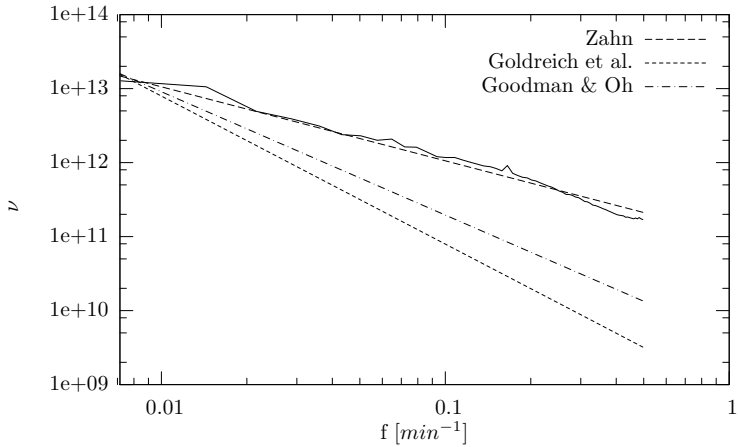}
\end{center}
\caption{Turbulent viscosity acting on a tidal flow in a stellar convection zone. The vertical component of that viscosity was determined by Penev et al. (2007) by applying an oscillating large-scale shear on a numerical simulation of turbulent convection; it decreases with the forcing frequency $f$. The result (in solid line) is compared here with several prescriptions that have been proposed for the loss of efficiency of turbulent friction when the tidal period becomes shorter than the convective turn-over time.}
\label{penev}
\end{figure}

One anticipates that the question will eventually be settled through high resolution numerical simulations of turbulent convection. Shortly after this summer school, I learned that a first step had been taken by Penev et al. (2007), who studied the dissipation of an imposed large-scale shear flow, periodic in time with period  $P_{\rm forc}$, using a 3-D convection code. They followed the method outlined by Goodman and Oh (1997) to derive the viscous stress  tensor. They confirmed that convection acts indeed as a turbulent viscosity on such a flow, since the off-diagonal components of the viscous tensor are much smaller than its diagonal components. They also observed that the vertical component of that tensor is about twice that of the horizontal components, which reflects the asymmetry of turbulent convection. Moreover, as can be seen in Fig.~\ref{penev} reproduced from their article, they found that this turbulent viscosity decreases as $f^{-1}$, where $f=P_{\rm forc}^{-1}$ is the forcing frequency, when the forcing period $P_{\rm forc}$ becomes shorter than the convective turn-over time. Hence they validate the first recipe (\ref{nuz}) quoted above, although it remains to be seen whether their result carries into more turbulent regimes.

\subsection{Beyond the weak friction approximation}\label{reduc}
When the turbulent viscosity depends on the tidal period, as in the prescriptions presented above, the weak friction approximation no longer applies. It is then necessary to break the tidal potential in its multiple Fourier components, and to sum up the torques exerted by each of these. Keeping only the second spherical harmonics of the potential, which is sufficient for most purposes, one has
\begin{equation}
U = {q \over 4} \omega^2 r^2 P_2^2(\cos \theta) \sum_l c_l \cos(2 \phi - l \omega t),
\label{U-fourier}
\end{equation}
and a similar expression for the axisymmetrical part in $P_2(\cos \theta)$. (More general expressions are given by G. Savonije in this volume.) The coefficients $c_l$ are functions of the eccentricity $e$;   to second order in $e$, for example, these are
\begin{equation}
c_0 = 0, \qquad c_1= -{1 \over 2}e, \qquad c_2=1-{5 \over 2} e^2, \qquad c_3 = {7 \over 2} e, \qquad c_4={17 \over 2} e^2 .
\end{equation}
In the frame of the rotating star, each of these components of the tidal potential produces a tidal flow of frequency $[m\Omega -  l\omega]$, which experiences a different turbulent viscosity $\nu_t$, if one takes into account the loss of efficiency when the tidal period becomes shorter than the convective turnover time (cf. \S\ref{fast-tide}). This is reflected in the coefficient $\lambda_2$ introduced above in (\ref{lambda2}), which takes a different value $\lambda^{m,l}$ for each tidal frequency. In a star with an outer convection zone, such as a late-type main-sequence star or a red giant, this parameter varies approximately as
\begin{equation}
\lambda^{m,l} = 0.019 \, \alpha^{4/3} \left( {3160 \over 3160 + \eta^2} \right)^{1/2} \quad \hbox{with} \;
\eta = [ m\Omega - l\omega] \, t_{\rm conv} ,
\end{equation}
where $ t_{\rm conv}$ is given by (\ref{tconv}) and $\alpha$ is the familiar mixing-length constant.

The equations governing the orbital evolution of the binary system then take the following form, to lowest order in $e$ (Zahn 1989):
\begin{eqnarray}
{d \ln a \over dt} \!\!\!\!\!\!\!\!\!&&=  - {12 \over  t_{\rm conv}} \, q(1+q)   \left({R \over a}\right)^8 
\left( \lambda^{2,2}  \left[1 - {\Omega \over \omega} \right]  \right. \\
&& + \,e^2  \left. \left\{ {3 \over 8} \lambda^{0,1}
+ {1 \over 16} \lambda^{2,1} \left[1 - 2{\Omega \over \omega} \right]
- 5 \lambda^{2,2}  \left[1 - {\Omega \over \omega} \right]
+ {147 \over 16} \lambda^{3,2} \left[3 - 2  {\Omega \over \omega} \right] \right\} \right), \nonumber
\label{orb1}
\end{eqnarray}
\begin{eqnarray}
{d \ln e \over dt} \!\!&=& \!\! - {3 \over  t_{\rm conv}} \, q(1+q)   \left({R \over a}\right)^8 \\
&&\times \left({3 \over 4} \lambda^{0,1}
- {1 \over 8} \lambda^{2,1} \left[1 - 2{\Omega \over \omega} \right]
- \lambda^{2,2}  \left[1 - {\Omega \over \omega} \right]
+ {49 \over 8} \lambda^{2,3} \left[3 - 2  {\Omega \over \omega} \right] \right), \nonumber
\label{circ1}
\end{eqnarray}
plus similar contributions of the secondary star. Note that we have added here the contribution of the axisymmetric part of the perturbing potential (which varies also in time when the orbit is eccentric, and yields the term in $ \lambda^{0,1}$). The angular velocity of the primary star obeys
\begin{eqnarray}
{d  \over dt} (I \Omega) \!\!&=& \!\! {6 \over  t_{\rm conv}} \, q^2 M R^2  
\left({R \over a}\right)^6 
\bigg( \lambda^{2,2} [\omega - \Omega]  \\
&& +   e^2  \left\{ {1 \over 8} \lambda^{2,1} [\omega - 2 \Omega]
- 5 \lambda^{2,2}  [\omega - \Omega] 
+ {49 \over 8} \lambda^{2,3} [3\omega - 2 \Omega] \bigg\} \right), \nonumber
\label{synchr1}
\end{eqnarray}
and likewise for the secondary star.
One verifies that (\ref{synchr1}) reduces to (\ref{sync}) and (\ref{circ1}) to (\ref{circ}) when all $\lambda^{m,l} \rightarrow \lambda_2$, in the weak friction approximation. 

These equations have been established assuming that all spins are aligned, i.e. that the rotation axis are orthogonal to the orbital plane.

\subsection{The quality factor}
Those working in planetary sciences often prefer to characterize the tidal dissipation by a dimensionless quality factor $Q$ defined as
\begin{equation}
Q^{-1} = {1 \over 2 \pi E_0} \oint \left(-{d E \over dt} \right) dt ,
\label{qprime}
\end{equation}
where $E_0$ is the maximum energy associated with the tidal distortion and the integral is the energy lost during one complete cycle (Goldreich \& Soter 1966). 

When applied to solid, elastic planets, this $Q$ is always combined with the Love number $k_2$, which measures the mass concentration in the star; it is then convenient to introduce $Q' = 3 Q / 2 k_2$, which reduces to $Q$ for a homogeneous body (Ogilvie \& Lin 2007). In fluid bodies, such as stars with convection zones or giant planets, the tidal torque is given by an integral over the star of the turbulent viscosity, as we have seen above in \S\ref{turbconv}, and $Q'$ is related to the coefficient $\lambda_2$ we have introduced there:
\begin{equation}
{1 \over Q'} = {8 \over 3} {\lambda_2 \over t_{\rm conv}} \left( { R^3 \over G M} \right) (\omega - \Omega) .
\label{qprime1}
\end{equation}
Usually $Q'$ is treated as a positive quantity, and the sign of the tidal  torque is imposed according to that of $(\omega - \Omega)$.

We see that $Q'$ depends both on intrinsic properties of the star (or the planet) and on the degree of synchronism, and this fact is often overlooked when comparing the $Q'$ of different planets or satellites in the solar system. 
If, as it has been suggested (cf. Ogilvie \& Lin 2007), the circularization period of late-type binary stars is roughly consistent with $Q'=10^6$, it would mean that $\lambda_2$ is inversely proportional to the tidal frequency $(\omega - \Omega)$, hence that the turbulent viscosity is reduced according to the first prescription (\ref{nuz}). 
On the other hand, if one chooses the quadratic reduction (\ref{nugn}), as they do in their paper, $Q'$ should scale as the tidal frequency.

\section{Comparing with the observations the theory of the equilibrium tide}
Having identified the most efficient dissipation mechanism, namely turbulent convection acting on the equilibrium tide, we shall now examine how well it accounts for the observed properties in binaries involving at least one component possessing an outer convection zone. We shall treat in turn the case of solar-type binaries on the main-sequence, that of such binaries during their pre-main sequence phase, and finally that of binaries in which one component has evolved to the giant stage.

\subsection{Solar-type binaries on the main sequence}
Applying eq. (\ref{circ}) to a binary of equal components, and of age $t_{\rm age}$, one finds that its orbit should be circular if its period is less than about
\begin{equation}
P_{\rm circ} =  6 \left({t_{\rm age} \over 5 \,  {\rm Gyrs} } \right)^{3/16}  {\rm days} .
\end{equation}
To obtain this result we assume that the rotation is synchronized with the orbital motion, and that the eccentricity decreased from $e=0.30$, a typical value for non-circularized binaries, to $e=0.02$, taken as detection threshold for the circularization. (If one takes this threshold to be $e=0.05$, the circularization period increases to 6.92 days for 5 Gyrs.)

Koch and Hrivnak (1981) were the first to compare this theoretical prediction with the distribution $e(P)$ of field binaries drawn from Batten's catalogue, and they found them to be compatible, although the transition period $P_{\rm circ}$ between circular and elliptic orbits was rather poorly defined, as one may expect with such a heterogeneous sample of stars mixing different ages.
\begin{figure}[!ht]
\begin{center}
\includegraphics[width=0.7\textwidth] {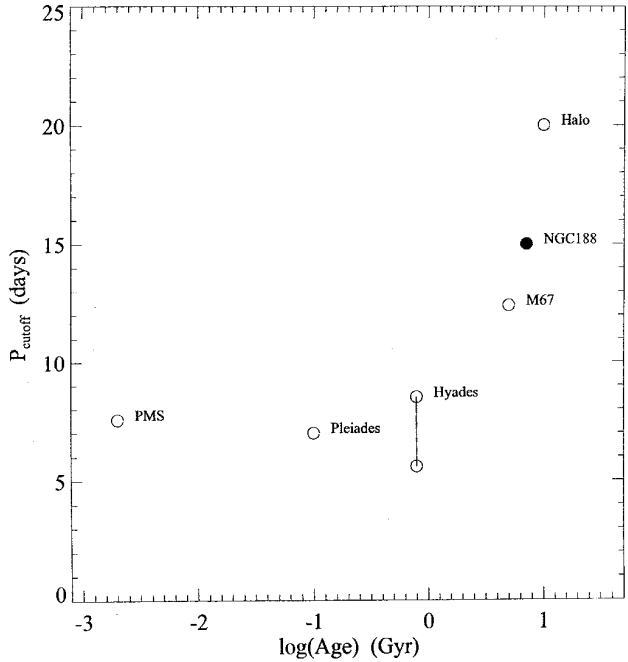}
\end{center}
\caption{Transition periods for circularization vs. age for six coeval stellar samples: PMS (Melo et al. 2001), Pleiades (Mermilliod et al. 1992), Hyades (Duquennoy et al. 1992), M67 (Latham et al. 1992), NGC 188 (Mathieu et al. 2004) and Galactic halo stars (Latham et al. 1992).  Note the near constancy of this period below 1 Gyr, at about $P_{\rm circ} \approx 8$ days, and its increase with age beyond. (From Mathieu et al. 2004; courtesy ApJ.)}
\label{cluster-eP}
\end{figure}

But the fact that the transition period is a slowly increasing function of age should be observable, by measuring the eccentricity of coeval cluster binaries.
Such a trend was found indeed by comparing the results of several surveys (Mermilliod et al. 1992; Duquennoy et al. 1992; Latham et al. 1992).
 This led Mathieu and Mazeh (1988) to suggest that the determination of $P_{\rm circ}$ could serve to evaluate the age of a cluster. However for M67, a cluster of about solar age, they found that the transition period was located between 10.3 and 11 days, well above the predicted 6 days, suggesting that tidal dissipation was about 20 times more efficient than inferred from the mixing-length theory.

A recent update was made by Mathieu et al. (2004); it summarizes the beautiful work accomplished over more than a decade by several dedicated teams (Fig.~\ref{cluster-eP}). As discussed by T. Mazeh in this volume, the transition period for circularization increases with age beyond 1 Gyr, but it is more or less constant below, around $P_{\rm circ} \approx 7-8$ days. It thus appears that two different mechanisms are at work, one operating on old binaries, and another that circularizes the young binaries. In fact, the latter had already been identified some time before.

\subsection{Circularization during the pre-main-sequence phase}
\label{pms-circ}
The orbital circularization depends strongly on the radius of the star: according to (\ref{circ}) $-d \ln e/ dt \propto R^8$. One thus expects that most of this circularization should occur on the PMS, where the stellar radius is much larger than on the main-sequence. Following this remark first made by Mayor and Mermilliod (1984), I undertook with L. Bouchet to integrate the equations describing the tidal evolution of solar-type binaries, starting from the birthline defined by Stahler (1983, 1988). Since on the PMS the convective turnover time can exceed the orbital period, it will also exceed the period of most Fourier components present in the tidal perturbation (cf. \ref{U-fourier}), and therefore one must take into account the reduction of the turbulent viscosity, as discussed in \S(\ref{reduc}).
\begin{figure}[!ht]
\begin{center}
\includegraphics[width=0.7\textwidth] {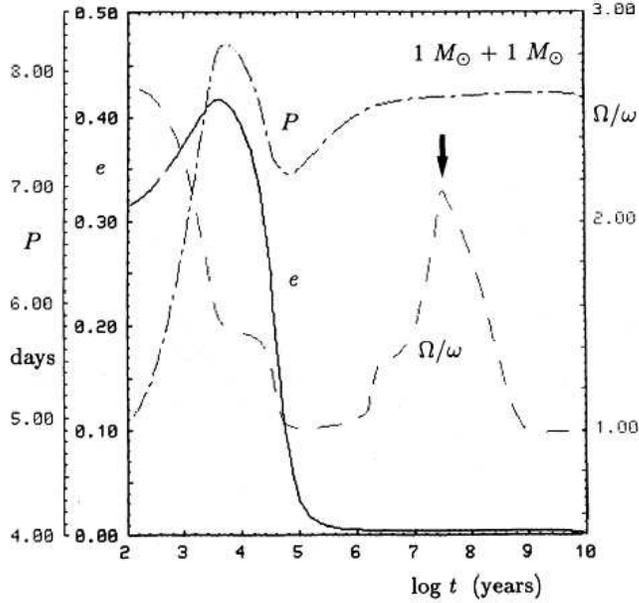}
\end{center}
\caption{Evolution in time of the eccentricity $e$, the orbital period $P$ and of the ratio between rotational and orbital frequencies $(\Omega / \omega)$, for a system with two components of 1~M$_\odot$. The initial period has been chosen such that the eccentricity would decrease from 0.300 to 0.005 when the binary reaches the zero age main-sequence (indicated by an arrow). (From Zahn \&  Bouchet 1989; courtesy A\&A.)}
\label{pms}
\end{figure}

The result is depicted in Fig. \ref{pms}, for a binary consisting of two solar-mass stars. The initial conditions were taken as $R=4.79 R_\odot$, $e=0.3$, 
$(\Omega / \omega)=3$, and the orbital period $P$ was chosen such that the eccentricity would drop to 0.005 when the binary reaches the zero age main-sequence (ZAMS). The rotation quickly synchronizes with the orbital motion  (in less than $10^5$ yrs), but thereafter the tidal torque weakens because the convection zone retreats, while the star keeps contracting; therefore the rotation speeds up again to about $(\Omega / \omega)=2$ at the ZAMS. Once the star  has settled on the MS, synchronization proceeds unhindered, and is achieved by an age of 1 Gyr.   The eccentricity first increases, as long as $(\Omega / \omega)>18/11$ (cf. Eq.~\ref{circ}), and then it steadily decreases to reach its final value $e=0.005$ at the ZAMS. Little circularization occurs thereafter on the MS. Angular momentum is transferred from the rotation to the orbit, which explains why the orbital period increases from 5 to 7.8 days. This final period depends rather weakly on the mass of the components, and it represents thus the transition period for circularization, in the absence of other tidal braking mechanisms.

This transition period agrees remarkably well with the properties of late type binaries younger than 1 Gyr, including the PMS stars, and thus there is little doubt that the circularization in these stars is due to the action of the equilibrium tide early on the PMS. The main uncertainties in the theoretical prediction are the initial radius $R_i$ ($P_{\rm circ}$ scales as $R_i$ to the power 15/16) and the prescription used to reduce the turbulent viscosity when the tidal period becomes shorter than the convective turnover time. We took here the linear prescription (\ref{nuz}); with the other, quadratic prescription (\ref{nugn}) the predicted transition period would be substantially shorter, contrary to what is observed.

It is important to note that binaries in their early MS stage may be circularized while not synchronized, which may seem paradoxical since the synchronization time (\ref{sync}) is much shorter than the circularization time (\ref{circ}). It stresses the necessity of following the whole tidal evolution of a given binary, starting from `reasonable' initial conditions.

\subsection{Circularization of binaries evolving off the main-sequence}
Another very interesting test for the tidal theory was performed by Verbunt and Phinney (1995), who chose for that a sample of wide binaries containing a giant star, because they wished to avoid what they call the ``troublesome problem of pre-main sequence circularization''  we  just discussed. Moreover,  in such binaries the tidal period exceeds the convective turnover time, so  that there is no need to worry about reducing the turbulent viscosity.

\begin{figure}[!t]
\begin{center}
\includegraphics[width=0.65\textwidth] {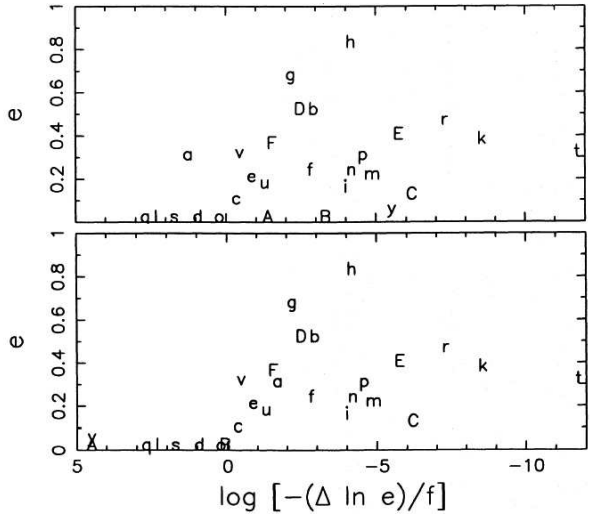}
\end{center}
\vskip -6pt
\caption{Observed eccentricities of binaries including a giant component vs. the change in eccentricity predicted by the tidal theory, invoking the equilibrium tide with turbulent dissipation in the convection zone (Verbunt \& Phinney 1995; courtesy A\&A). In the upper panel the giant component is assumed to be on the asymptotic giant branch; some corrections have been applied to obtain the result of the lower panel (see text).}
\label{vb95}
\end{figure}

They selected 29 binaries with giant components in several galactic clusters, whose age and distance are well established. 
They integrated the circularization equation (\ref{circ}) for these binaries from the MS to  their present location in the HR diagram, and presented the result in the form $-\Delta \ln e /f$, where $\Delta \ln e$ is the cumulated change in eccentricity, and $f$ a factor that depends on the convection theory used to calculate the turbulent dissipation. For the classical mixing-length treatment that was employed in \S\ref{turbconv}, $f = (M / M_{\rm env})^{2/3} (\alpha/2)^{4/3}$,  $M_{\rm env}$ being the mass of the convective envelope and $\alpha$ the mixing-length parameter, and therefore according to theory $f$ should be of order unity\footnote{In their eq. 1, Verbunt \& Phinney parametrize our coefficient $\lambda_2$ (cf.  Zahn 1989) in terms of the envelope mass, which explains why the expression of $f$ quoted here differs somewhat from their's, with little consequence when $M_{\rm env} \rightarrow M$.}.

Fig. \ref{vb95} displays the observed eccentricity of these binaries (each individually labelled by a letter) as a function of the predicted drop in eccentricity $-\Delta \ln e$, or rather $\log [-\Delta \ln e /f]$. For $\Delta \ln e /f > 1$ (i.e. $-\log [-\Delta \ln e /f] < 0$), the orbit should be circularized, whereas it should not for $-\log [-\Delta \ln e /f] > 0$. Phinney and Verbunt assume that all binaries are presently on the asymptotic giant branch (core helium burning), because they stay there 10 times longer than previously on the red giant branch (shell hydrogen burning).

The result is shown in the upper panel: the great majority of binaries complies with the theoretical prediction, displaying circular orbits for $-\log [-\Delta \ln e /f] > 0$ and eccentric obits for $-\log [-\Delta \ln e /f] < 0$.
However there are 4 notable exceptions: binary `a' has kept an eccentricity of 0.30, while its orbit should be circular, and binaries `A', `B',  `y'  have circular orbits, where these should be elliptic. Phinney and Verbunt concluded that binary `a' must still be ascending the red giant branch, and that the other 3 binaries must have undergone an exchange of matter, which very efficiently circularizes the orbit, and therefore that they should have an evolved companion, such as a white dwarf. After these adjustments, the 4 binaries are no longer exceptions, as can be seen in the lower panel; moreover,  the fact that the transition from circular to elliptic orbits occurs in the vicinity of  $\log [-\Delta \ln e /f] \approx 0$ confirms that the parameter $f$ is indeed of order unity, thus validating the theory of the equilibrium tide with turbulent dissipation.

Two years later Landsman et al. (1997) announced that the secondary of S1040 in M67, the binary labelled `A', is indeed a white dwarf, confirming the brilliant conjecture of Verbunt and Phinney that it must have experienced an episode of mass exchange.

We may thus conclude that turbulent viscosity acting on the equilibrium tide explains most observations, with the important exception of the circularization of main-sequence binaries older than about 1 Gyr, for which it seems that we have to seek another dissipation mechanism. A very plausible candidate for that is the dynamical tide, which we shall examine next.

\section{The dynamical tide}
Due to its elastic properties, a star can oscillate in various modes: acoustic modes, internal gravity modes, inertial modes, where the restoring force is respectively the compressibility of the gas, the buoyancy force in stably stratified regions, and the Coriolis force in the rotating star. If their frequency is low enough, these modes can be excited by the periodic tidal potential; the response is called the {\it dynamical tide}. 

\begin{figure}[!ht]
\begin{center}
\includegraphics[width=0.6\textwidth,angle=1] {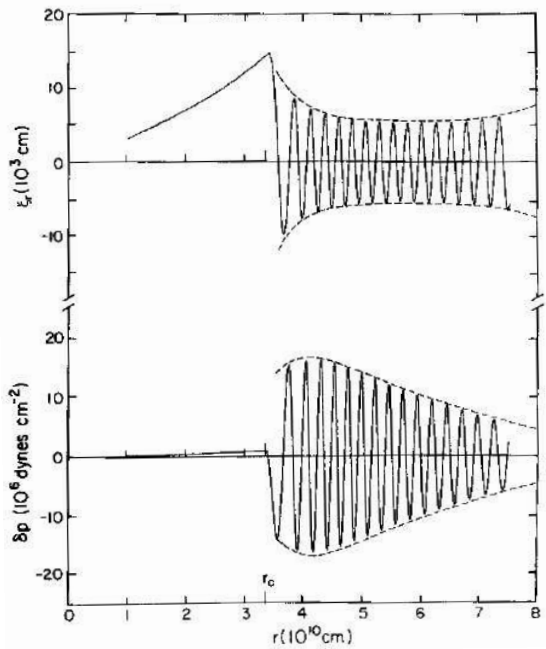}
\end{center}
\caption{Gravity mode in a 5 M$\odot$ main sequence star excited by a tidal potential of frequency $\sigma=2 \, 10^{-5}$ s$^{-1}$: $\xi_r$ is the radial displacement, and $\delta p$ the pressure perturbation. The radius of the star is $1.88\,10^{11}$ cm; only the inner region is shown here  (from Goldreich \& Nicholson 1989; courtesy ApJ).}
\label{grav}
\end{figure}

\subsection{Gravity modes excited by a tidal potential}\label{grav-exc}
The modes that have received most attention so far are the tidally excited gravity modes; associated with radiative damping, they have first been invoked for the tidal evolution of massive main-sequence binaries (Zahn 1975). Their restoring force is provided by the buoyancy, whose strength is measured by the buoyancy frequency $N$, which is given by
\begin{equation}
N^2 = {g  \delta \over H_P}\left[ \left({\partial \ln T \over \partial \ln P}\right)_{\rm \! ad} -{d \ln T \over d \ln P} + 
{\varphi \over \delta} \, {d \ln \mu \over d \ln P}\right],
\end{equation}
using classical notations, and $\mu$ being the molecular weight ($\delta= -( \partial \ln \rho / \partial \ln P)_{T, \mu}$ and $\varphi=(\partial \ln \rho / \partial \ln \mu)_{T, P}$ are unity for perfect gas).
 
The modes that are most excited are those whose frequency is close to the tidal frequency, and these are of high radial order: typically they have more than 10 or 20 radial  nodes in the radiation zone, because their wavelength scales as $\lambda_r  \propto r \sigma /  N$, and because the tidal frequency $\sigma$, of the order of  days$^{-1}$, is much lower than the buoyancy frequency $N$, of the order of 1 hour$^{-1}$. See Fig.~\ref{grav} for a typical example of such modes, in a 5 M$_\odot$ ZAMS star. Dissipation has been neglected, and therefore the mode is an adiabatic standing wave; note that it is evanescent in the convective core. 

These gravity modes couple with the periodic tidal potential in the vicinity of the convective core, whereas their damping occurs mainly near the surface, because the thermal damping time, which scales roughly as the cube of the temperature,  is much shorter there than in the deep interior. The angular momentum drawn from the orbit is deposited near the surface, and hence it is the surface layers that are synchronized first with the orbital motion.  As was emphasized by Goldreich and Nicholson (1989),  this synchronization is further sped up because the local tidal frequency experienced by the fluid entrained in the differential rotation, $\sigma = 2\Omega(r)-2\omega$, tends to zero, and so does also the radial wavelength $\lambda_r$, as we have seen above, thus enhancing the damping.  
\begin{figure}[!t]
\begin{center}
\includegraphics[width=0.8\textwidth] {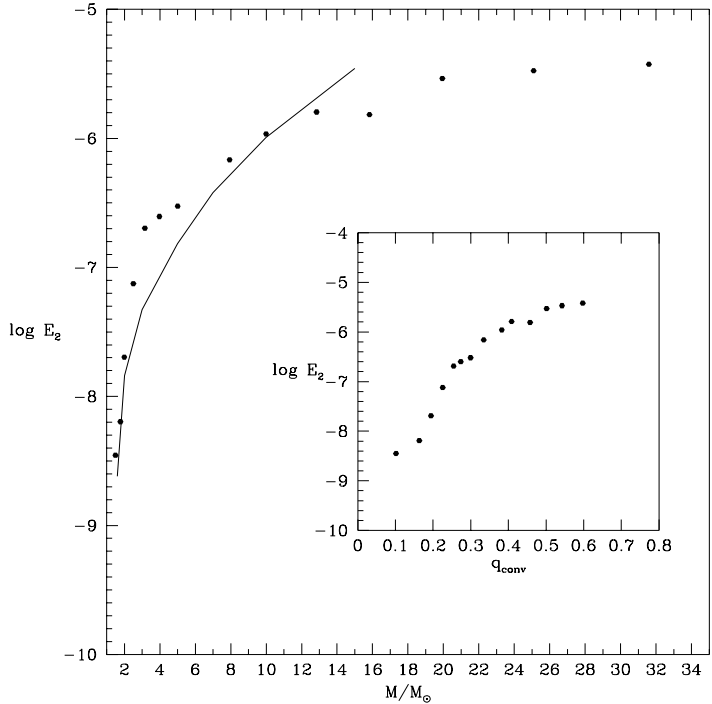}
\end{center}
\caption{Tidal parameter $E_2$ characterizing the strength of the dynamical tide,
cf. (\ref{sync-mass}) and (\ref{circ-mass}). It is displayed here on a logarithmic scale, as a function of mass (in solar units), near the ZAMS. The solid line reports  the results of former computations by Zahn (1975). The insert shows the dependence of $E_2$ on the relative size of the convective core (from Giuricin et al. 1984; courtesy A\&A).}
\label{clar-cunh}
\end{figure}

For low enough tidal frequency, the tidal wave is completely damped (meaning that is has become a pure propagating wave),
and one can use the WKB treatment to evaluate the total torque applied on the star (Zahn 1975). For the synchronization time (assuming uniform rotation) one finds 
\begin{equation}
{1\over t_{\rm sync}} = - {d \over dt} \left|{2(\Omega - \omega) \over
\omega}\right|^{-5/3} =
5 \left({G M \over R^3}\right)^{1/2} q^2 (1+q)^{5/6}\, {M R^2 \over I}
E_2 \left({R \over a}\right)^{17/2} ,
\label{sync-mass}
\end{equation}
and likewise for the circularization time, assuming that synchronization has been quickly achieved:
\begin{equation}
{1\over t_{\rm circ}} = - {d \ln e \over dt}  =
{21 \over 2}  \left({G M \over R^3}\right)^{1/2} q (1+q)^{11/6} \,
E_2 \left({R \over a}\right)^{21/2} ;
\label{circ-mass}
\end{equation}
the companion star contributes a similar amount.
$E_2$ is a parameter measuring the coupling between the tidal potential and the gravity mode: it depends sensitively on the size of the convective core, and thus on the mass of the star. Its expression is given in Zahn (1975); it has been tabulated by Claret and Cunha (1997) for various stellar models, as shown in Fig.~\ref{clar-cunh} ; for a 10 M$_\odot$ ZAMS star, it is $E_2\approx 10^{-6}$.

This theory was initially developed for pure gravity modes, and as such it was strictly applicable only to non-rotating stars. It was later extended  by Rocca (1989)  to (uniformly) rotating stars; she showed that taking the Coriolis force into account modifies only slightly the results presented above.

\begin{figure}[!ht]
\begin{center}
\includegraphics[width=\textwidth] {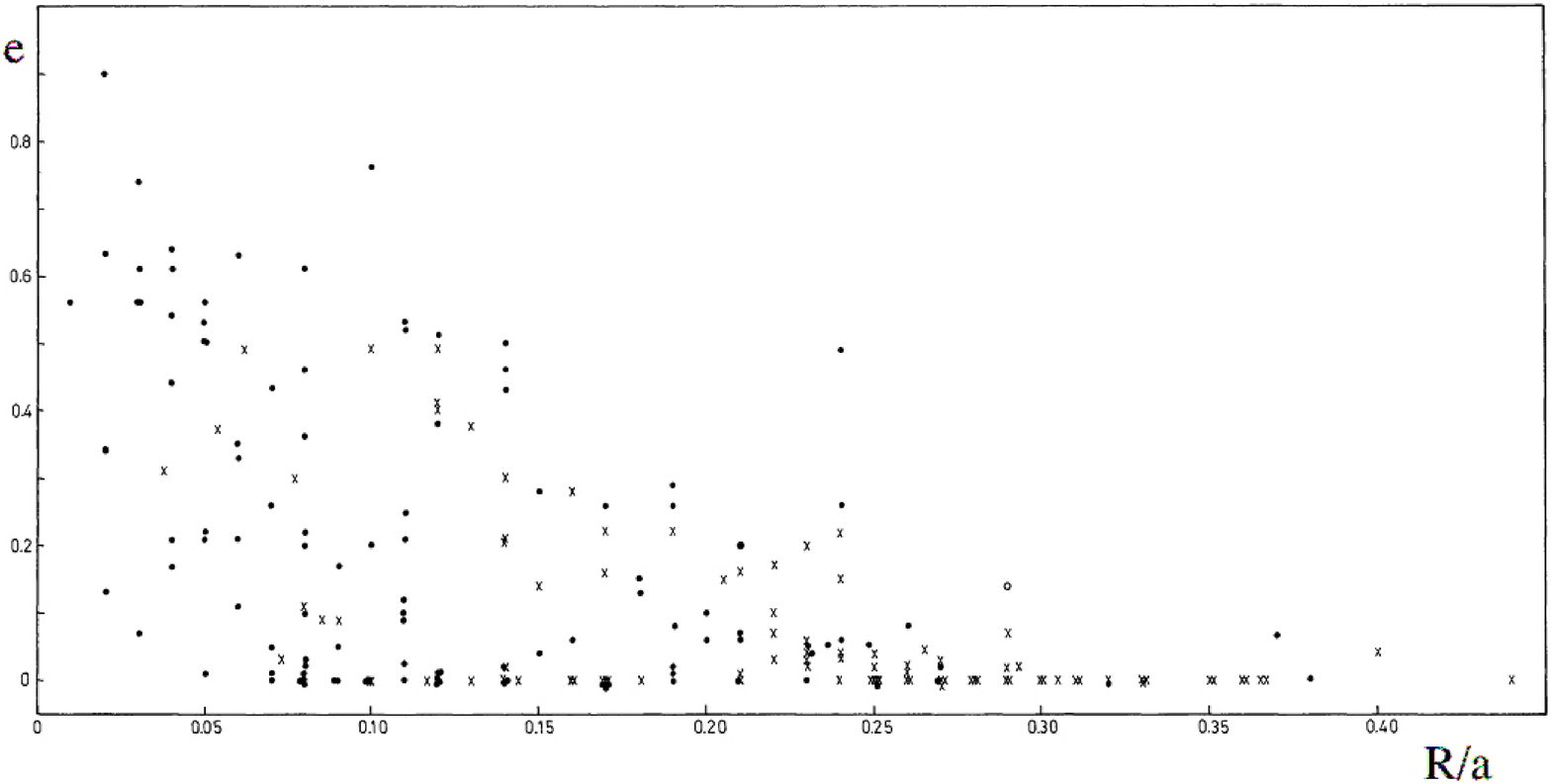}
\end{center}
\caption{Eccentricity $e$ vs. fractional radius $R/a$ for early-type binaries (spectral types O, B, F) listed in Batten's catalogue (from Giuricin et al. 1984; courtesy A\&A).}
\label{giur}
\end{figure}

\subsection{Circularization of early-type binaries}
Giuricin et al. (1984) were the first to compare the predictions of tidal theory with the properties of early-type binaries.  Applied to binaries with two identical components of mass between 2 and 15 M$_\odot$, eq. (\ref{circ-mass}) predicts a transition value of $R/a \approx 0.25$ for the fractional radius, i.e. the radius expressed in units of semi-major axis\footnote{This value depends little on mass (Zahn 1977); translated into tidal periods, the transition periods would spread between 1 to 2 days, which explains why it is preferable to use $R/a$ for the observational test.}. This value is in good agreement with the observed distribution of eccentricities vs. fractional radius displayed in Fig.~\ref{giur}, although many binaries are circular for $R/a < 0.25$.

\begin{figure}[!t]
\begin{center}
\includegraphics[width=\textwidth] {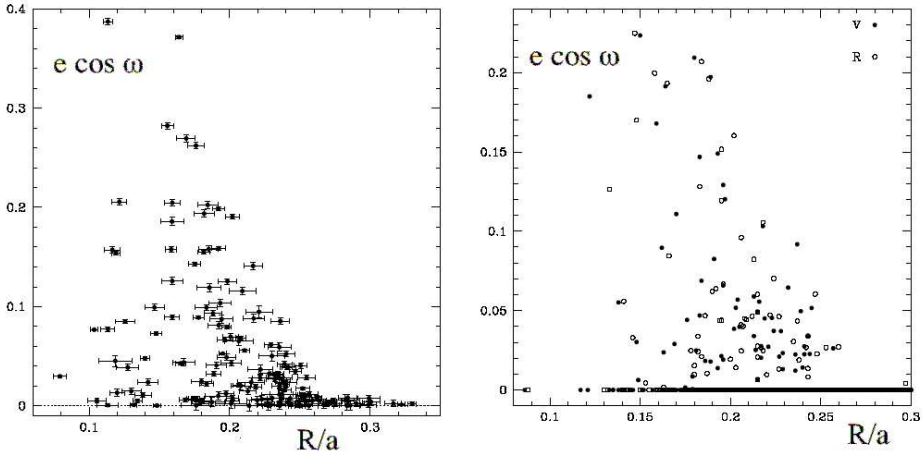}
\end{center}
\caption{ Left  panel: $e \cos \omega$ vs. relative radius $R/a$ for detached eclipsing binaries in the SMC.  Right panel: same for the LMC, full dots based on V lightcurves, open dots based on R lightcurves.  (From North \& Zahn 2003; courtesy A\&A.)}
\label{north}
\end{figure}

A similar investigation was recently carried out on eclipsing binaries which had been detected in the Magellanic Clouds during the MACHO and OGGLE campaigns (North \& Zahn 2003). The results are shown in Fig.~\ref{north}. Here again the $e$ vs. $R/a$ distribution strongly suggests a transition value of
$R/a=0.255$, in excellent agreement with theory. However  an important fraction of binaries are circular at lower fractional radius: it is as if there were two populations of binaries, one complying with the predictions above, and the other experiencing another, more efficient tidal damping. Histograms of the eccentricity distribution at given $R/a$ confirm that impression, and so does also a much wider survey carried out by Mazeh et al. (2006) (cf. his chapter in this volume).

On may wonder why the binaries in the Magellanic Clouds behave so similarly to those in our Galaxy: they have lower metallicities, and therefore somewhat larger convective cores, and one would expect that these differences be reflected in the coefficient $E_2$. However the radii differ too, and the two effects compensate each other such that the predicted transition periods are very nearly the same.

\subsection{Resonance locking in early-type binaries}
Recently Witte and Savonije (1999a, 1999b) revisited the theory of the dynamical tide, by making full account of the Coriolis force, while still neglecting the effect of the centrifugal force. Instead of projecting the forced oscillations on spherical functions, they solved the governing equations directly in two dimensions ($r, \theta$), for several values of the angular velocity $\Omega$ and of the tidal frequency $\sigma=l\omega-2\Omega$ in the rotating frame. 
 When the orbit is circular and the star rotates in the same sense as the orbital motion, only one retrograde mode can be excited at $ \sigma = 2\omega - 2 \Omega$.  But when the orbit is elliptic, many other tidal frequencies appear:
$ \sigma_l = l \omega - 2 \Omega$ with $|l|= 1, 3,$ etc. (see \S\ref{reduc}), and both retrograde and prograde modes can be excited. 
Therefore it is very likely that a binary undergoes some resonances during its evolution, both because the tidal frequency shifts in the course of synchronization, and because the eigenfrequencies are affected by the structural changes of the stars.

The effect of resonances on tidal evolution was largely ignored earlier (Zahn 1975; Rocca 1989; Goldreich \& Nicholson 1989) on the grounds that stars would move quickly through such resonances, because their width $\Delta \sigma$ is inversely proportional to their amplitude. But, most interestingly, Witte and Savonije (1999b) pointed out that this is not necessarily true, and that a binary can be trapped into a resonance, when one takes into account the whole set of tidal frequencies associated with an elliptic orbit.  Retrograde and prograde modes exert torques of opposite sign, and when they balance each other, they may lock the star into such resonances. Moreover, structural changes also can conspire to favor such locking. The consequence is that circularization is sped up by such resonances, as demonstrated by several specific cases they have studied, which are discussed in detail by G. Savonije in this volume. The results are rather sensitive to the initial conditions, which may explain the observations mentioned above concerning the Magellanic Clouds binaries, namely that for the same orbital period (or fractional radius), some binaries are circular while the others are not, as if there were two tidal damping mechanisms.

Recently Willems et al. (2003) too examined the behavior of a 5 M$_\odot$ binary in the vicinity of resonances, using the quasi-adiabatic approximation;  however they did not include the Coriolis force, which according to Witte and Savonije plays an important role in coupling the eigenmodes.

\subsection{Resonance locking in late-type binaries}
Let us come back to the late-type main-sequence binaries. We have seen that turbulent dissipation of the equilibrium tide, at least in its present state, cannot explain the circularization observed in binaries older than 1 Gyr. This incited Terquem et al. (1998) and  Goodman and Dikson (1998), to examine whether the dynamical tide could not be responsible for the observed circularization.  Both teams invoked radiative damping as dissipation mechanism, as had been done previously for early-type stars. But here such damping is rather weak, because the oscillation modes are evanescent in the convection zone, where thermal dissipation would be strongest. Therefore, contrary to what has been found in early type stars, oscillations modes can enter in resonance at very low tidal frequency, i.e. very close to synchronization. This means that one has to deal with modes which have up to thousand radial nodes, which puts a serious burden on the numerical work, as experienced by Terquem et al.; they restricted their exploration to the vicinity of 3 orbital periods, but included turbulent dissipation in the convection zone, where the modes are evanescent. On the contrary, Goodman and Dikson chose a semi-analytical WKB approach, much as in Zahn (1977).

Though their quantitative results differ somewhat, the conclusions of the two teams agree, namely  that the dynamical tide  cannot account for the circularization of the oldest late-type binaries; comparing the predicted transition periods, one sees that it is less efficient than the equilibrium tide. 

The problem was re-examined shortly after by Witte and Savonije (2002), who anticipated that here also resonance locking could play an important role. They made account of the Coriolis force, but refrained from the direct 2D calculations they used for early-type binaries, which would be much more cumbersome given the high order of the modes. Instead, they retained only the radial component of the rotation vector, in the so-called ``traditional approximation'' (which is justified in the limit of low exciting frequency). The $r$ and $\theta$ variables then separate again, as in the non-rotating case; but the horizontal functions are the so-called Hough functions (Savonije \& Witte 2002), which depend on the rotation rate. 

Today this process of resonance locking in the dynamical tide thus appears as the most efficient, {\it on the main-sequence},  among all dissipation mechanisms that have been explored, as discussed by G. Savonije in this volume. When starting with quasi-synchronous or super-synchronous stars, the predicted transition period is a slowly increasing function of age; for $5\,10^9$ yrs, this period is about 7 days, thus higher than that predicted by the equilibrium tide (6 days). But even so, the theoretical predictions are well below the observed ones, unless one allows for very slow, and rather unrealistic initial rotation (period of 100 days). 
Let us recall that  below 1 Gyr the observations agree very well with the transition period derived for the PMS circularization through the equilibrium tide,  as we have seen in  \S\ref{pms-circ}

\section{Tidal damping through inertial modes}
While gravity modes propagate only in stably stratified regions, there is another type of modes, the inertial modes,  that are able to propagate also in neutrally stratified convection zones. 
They owe their existence to the Coriolis force, and hence their frequency, in the frame of the rotating star, is bound by $2 \Omega$. They may thus be excited by the tidal potential, much as the gravity modes,  provided the tidal frequency is less than the inertial frequency $2 \Omega$. 
These modes have received little attention so far, until very recently. Their properties are described in detail in this volume by M. Rieutord. 

\begin{figure}[!ht]
\begin{center}
\includegraphics[width=\textwidth] {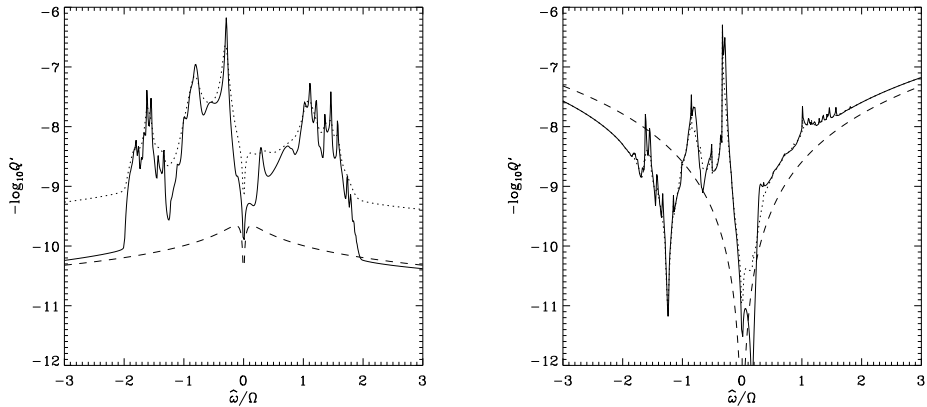}
\end{center}
\caption{ Dissipation rate $Q'$, defined in  (\ref{qprime}) and (\ref{qprime1}), as a function of the tidal frequency $\widehat\omega$ normalized by the rotation frequency $ \Omega$. The solar model has a spin period of 10d. Left: $Q'$ from the viscous dissipation of inertial modes in the convection zone. Right: $Q'$ from the excitation of Hough modes in the radiative zone. The dashed lines show the effect of omitting the Coriolis force, hence reducing the dissipation to that of the equilibrium tide. The turbulent viscosity has been reduced according to prescription (\ref{nugn}). The dotted lines show the result of increasing that turbulent viscosity by a factor of 10.
(From Ogilvie \& Lin 2007; courtesy ApJ.)}
\label{ogilin}
\end{figure}

Recently Ogilvie and Lin (2007) have studied numerically the r\^ole of these inertial modes in damping the tides, choosing a solar-type star. The results are depicted in Fig.~\ref{ogilin}. One sees that their contribution (left panel), through their viscous dissipation in the convection zone, can be as large as that of the gravito-inertial modes in the radiation zone (right panel). The dashed lines show the effect of switching off the Coriolis force, and the dotted line, in the left panel, that of increasing the turbulent viscosity by a factor 10. Note 
that Ogilvie and Lin opted for the quadratic reduction of that turbulent viscosity (eq. \ref{nuz}), which probably underestimates the contribution of the equilibrium tide.

A remarkable property of these inertial modes is that their peak amplitude, at resonance,
does not depend on the strength of the viscosity, as can be seen in the left panel of Fig.~\ref{ogilin}. This is because these modes are described  in the inviscid limit by an equation that is spatially hyperbolic, and hence their characteristic rays are focused on wave attractors, where most of viscous dissipation occurs, and whose thickness scales in such a way as to render the dissipation independent of viscosity. This is explained in detail by Ogilvie \& Lin, and by M. Rieutord in this volume.

\section{Conclusion and perspectives}
Let me summarize. The two tidal dissipation processes that have received most attention so far are the turbulent friction acting on the equilibrium tide, which was first described in the 60's (Zahn 1966b), and the radiative damping of the dynamical tide,  identified in the 70's (Zahn 1975). These processes operate  respectively in convection and radiation zones, and they have been quite successful in explaining the observed orbital circularization of binary stars. This is particularly true for the early-type MS binaries, for which we have now at our disposal very large samples gathered during the OGGLE and MACHO campaigns: their transition period is precisely defined and it agrees extremely well with that predicted by the theory of the dynamical tide, which is thus validated. However many of these binaries are circularized well above this transition period, as if they had experienced another, more efficient tidal dissipation mechanism. A very likely explanation for this behavior is that these binaries have undergone several episodes of resonance locking, as was described by Witte and Savonije (1999a, 1999b).

On the other hand, the equilibrium tide damped by turbulent dissipation accounts very well for the properties of binaries containing a red giant, as was demonstrated by Verbunt \& Phinney (1995). It also explains the transition period of about 8 days observed in late-type binaries that are younger than about 1 Gyr: the explanation is that these have been circularized during the PMS phase, when they were much more voluminous and fully convective.
The only serious discrepancy today seems to be the behavior of late-type main-sequence binaries older than 1~Gyr, whose transition period increases with age and is higher than that predicted when applying straightforward the theory of the equilibrium tide. Here again one may invoke
 the dynamical tide with resonance locking in the radiative core of these stars, as was shown by Witte and Savonije (2002).

Their mechanism appears thus highly promising, and it ought to be further explored. For instance, one should take into account that the tidal torque is applied primarily  to specific regions: the outer convection zone in late-type MS stars and the outermost part of the radiation zone in early-type stars.  These regions are synchronized more quickly then the rest of the star, and therefore differential rotation develops in their radiation zone. This increases the radiative damping, since the local tidal frequency tends then to zero as the tidal wave approaches the synchronized region, as we explained in \S\ref{grav-exc}.

For late-type binaries, a highly interesting alternative is offered by the damping of inertial waves in their convective envelope, which is being explored by Ogilvie and Lin (2007). This process is likely to play an important role also in giant planets (Ogilvie \& Lin 2004).
The difficulty in studying these waves is that they require highly resolved 2D numerical calculations, since the so-called traditional approximation is no longer applicable to render the problem 
separable.

Work is in progress on several other points, and I shall quote only a few.  Kumar and Goodman (1996) have studied the enhanced damping of the oscillations triggered in tidal-capture binaries, due to non-linear coupling between the eigenmodes, which is extremely strong in such highly eccentric orbits. Rieutord (2004, and in this volume) is examining the possibility that the so-called elliptic instability may occur in binary stars; this instability is observed in the laboratory when fluid is forced to rotate between boundaries that have a slight ellipticity, and it leads to turbulence. Even the equilibrium tide is being revisited, taking into account the differential rotation of the convection zone (Mathis \& Zahn, in preparation). 

Finally, it remains to explain why I made no attempt here to reconcile the theoretical predictions for the synchronization of the binary components with their observed surface rotation. The reason is that in most cases the tidal torque is applied mainly to the outermost part of the star, which is synchronized much more rapidly than the interior; therefore the interpretation of the surface rotation requires to model the transport of angular momentum within the star, and in particular where it proceeds slowest, i.e. in the radiation zones. 
This is a difficult task that only now begins  to be undertaken seriously:
for recent accounts on this problem, see the reviews by Talon (2007) and Zahn (2007). I am confident that we will see much progress in solving this problem when the next school will be held on that theme, hopefully in a not too distant future!

{}

\end{document}